# Metastable phase of UTe$_2$ formed under high pressure above 5 GPa

L. Q. Huston[1], D. Y. Popov[2], A. Weiland[1], M. M. Bordelon[1], P. F. S. Rosa[1], R. L. Rowland II[1], B. L. Scott[1], G. Shen[2], C. Park[2], E. K. Moss[1], S. M. Thomas[1], J. D. Thompson[1], B. T. Sturtevant[1], E. D. Bauer[1,a]

[1]Los Alamos National Laboratory, Los Alamos, NM 87545, USA

[2]HPCAT, X-ray Sciences Division, Argonne National Laboratory, Lemont, IL, 60439, USA

a) corresponding author, edbauer@lanl.gov

**Abstract**

Uranium ditelluride (UTe$_2$) has attracted recent interest due to its unique superconducting properties, which include the potential for a topological odd-parity superconducting state. Recently, ac-calorimetry measurements under pressure indicate a change in the ground state of UTe$_2$ from superconducting to antiferromagnetic at 1.4 GPa. Here, we investigate the effect of pressure on the crystal structure of UTe$_2$ up to 25 GPa at room temperature using x-ray diffraction. We find that UTe$_2$, which at ambient conditions has an orthorhombic (*Immm*) structure, transforms to a body-centered tetragonal (*I4/mmm*) structure at 5 GPa in a quasi-hydrostatic neon (Ne) pressure transmitting medium. In the absence of a pressure-transmitting medium, this transformation occurs between 5 and 8 GPa. The data were fit with a third-order Birch-Murnaghan equation of state resulting in values of $B_0$=46.0±0.6 GPa, B'=9.3±0.5 (no pressure medium) and $B_0$=42.5±2.0 GPa, B'=9.3 (fixed) (neon pressure medium) for the *Immm* phase. For the *I4/mmm* phase, $B_0$=78.9±0.5 GPa and B'=4.2±0.1 (no pressure transmitting medium), and $B_0$=70.0±1.1 GPa and B'=4.1±0.2 (neon pressure medium). The high-pressure tetragonal phase is retained after decompression to ambient pressure, with approximately 30% remaining after 2 days. We argue that the observed phase transition into a higher symmetry structure at P~5 GPa (orthorhombic to tetragonal), is accompanied by an increase in the shortest distance between uranium atoms from 3.6 Å (orthorhombic) to 3.9 Å (tetragonal), which suggests localization of the 5f electrons, albeit with a 10.7% decrease in volume.

# Introduction

Odd-parity topological superconductivity has received widespread attention due to its potential application for fault-tolerant quantum computing [1]. The recently discovered superconductor uranium ditelluride (UTe$_2$) has been proposed to be an odd-parity superconductor with potential topological properties, such as Weyl superconductivity [2-4]. Kerr effect measurements on UTe$_2$ provide evidence for time-reversal symmetry breaking in the superconducting state [4], and chiral superconductivity has been suggested as the origin of in-gap states via scanning-tunneling spectroscopy measurements [5]. The very small decrease of the Knight shift via $^{125}$Te nuclear magnetic resonance (NMR) measurements also suggests odd-parity superconductivity [6].

UTe$_2$ has multiple superconducting phases [2, 7], both under applied pressure and in high magnetic field. At ambient temperature and pressure, UTe$_2$ crystallizes in an orthorhombic structure with an *Immm* space group (a=4.162 Å, b=6.128 Å, c=13.965 Å) [8, 9] (Fig. 1a). High-quality crystals of UTe$_2$ exhibit superconductivity around a critical temperature of $T_c$=2 K [10, 11, 11a]. In high magnetic field, superconductivity persists up to an upper critical field of $H_{c2}$=35 T for field applied along the b-axis [7]. When the magnetic field is rotated 23° from the *b* axis towards the *c* axis, reentrant superconductivity is observed above 40 T [2, 12, 13].

The complex temperature-pressure (T-P) phase diagram of UTe$_2$ has been determined from electrical resistivity and ac-calorimetry measurements up to 2.5 GPa [3, 14, 15] and is shown in Fig. 1 (b) [15]. A second superconducting phase emerges above 0.3 GPa and reaches a maximum value of $T_c$=3.0 K at P=1.2 GPa. Two magnetic transitions (with temperatures $T_{m1}$ and $T_{m2}$) arise for P>1.4 GPa at $T_{m2}$=3 K and $T_{m1}$=4.5 K (at P=1.5 GPa). These transitions are consistent with antiferromagnetism and display a small coexistence region with superconductivity between 1.4 and 1.5 GPa [15]. Recent theoretical investigations based on the orthorhombic D$_{2h}$ space group propose a variety of possible superconducting order parameters, including triplet B$_{3u}$ + iB$_{2u}$ [4] or mixed pairing B$_{3u}$ + A$_g$ under pressure [15, 16]. The upper critical field and other phase boundaries have been found to change with pressure and several studies have mapped out the pressure-temperature-magnetic field phase diagram of UTe$_2$ up to 3 GPa [15, 17-19]. Ran *et al.* [18] studied UTe$_2$ up to 2 GPa and magnetic fields up to 40 T in multiple directions and found that, along the b-axis, both the "reentrant" superconductivity and the field polarized phase occurred at lower magnetic fields as pressure is increased.

Given the multiple magnetic and superconducting phases of $UTe_2$ under pressure, it is important to study how the crystal structure of $UTe_2$ changes with pressure. In this study, the evolution of the crystal structure of $UTe_2$ is determined at room temperature up to 25 GPa using diamond anvil cells coupled with angle dispersive x-ray diffraction (XRD). A structural phase transition occurs in $UTe_2$ at pressures above 5 GPa (at room temperature) from the orthorhombic (*Immm*) phase to a high-pressure, body-centered tetragonal phase (space group *I4/mmm*). This higher symmetry *I4/mmm* phase was found to be present upon decompression to ambient pressure, with approximately 30% remaining after 2 days. The thermodynamic and transport properties of $UTe_2$ are reported after reversion to the orthorhombic *Immm* phase approximately 2-3 weeks after the material was decompressed to ambient pressure. No signature of superconductivity is found below 2 K in magnetic susceptibility, electrical resistivity, or specific heat measurements.

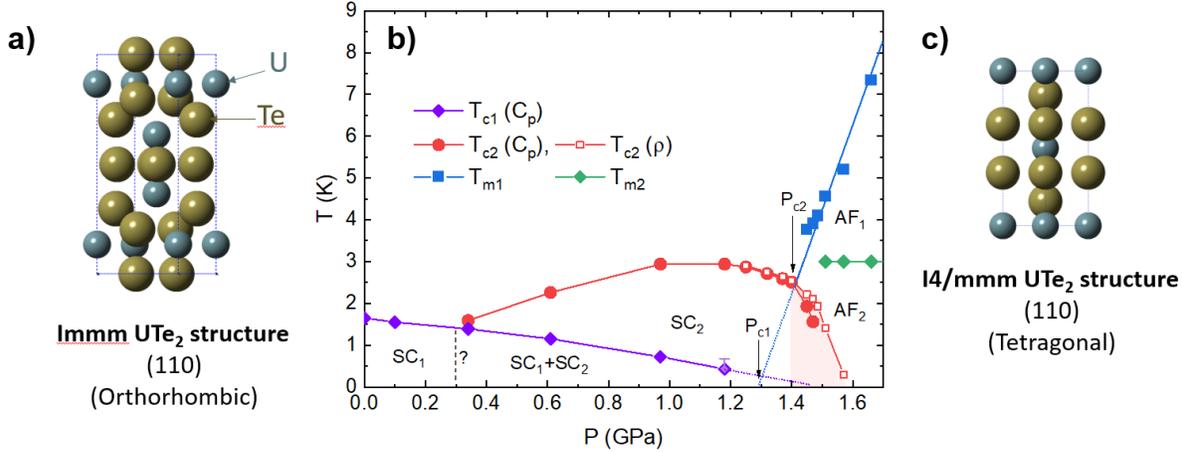

**Figure 1: a)** Orthorhombic (*Immm*) crystal structure of $UTe_2$ oriented along the (110) plane. U atoms are a teal color. **b)** Temperature-Pressure (T-P) phase diagram of $UTe_2$ up to 1.7 GPa. $SC_1$: ambient-pressure superconducting phase; SC2: high-pressure superconducting phase; $AF_1$, $AF_2$: high-pressure antiferromagnetic phases; $P_{c1}$: putative quantum critical pressure; $P_{c2}$: pressure where $T_{m1}$ is lower than $T_{c2}$. **c)** High-pressure tetragonal (*I4/mmm*) crystal structure of $UTe_2$ oriented along the (110) plane.

## Experimental Method

Superconducting single crystals of $UTe_2$ were grown using the chemical vapor transport method [7]. Several small crystals were crushed and passed through a 20 μm mesh sieve to collect the smallest particles. To make a polycrystalline sample, the sieved powder was pressed into a pellet using a die and press. Small pieces of the pellet were then broken off and loaded into a diamond anvil cell along with a Cu pressure standard and either no pressure transmitting medium (PTM) or a neon (Ne) PTM.

Five high pressure experiments were conducted using either BX-90 diamond anvil cells (DACs) [20] or symmetric DACs; both styles were modified to provide containment to prevent the release of the radioactive samples. Diamond culet diameters ranged from 150 to 400 μm. The gaskets used were either 40 μm thick $Co_{66}Si_{15}B_{14}Fe_4Ni_1$ metallic glass foil (Goodfellow), or 300 μm thick stainless steel pre-indented to a thickness of ~30 μm. The XRD experiments were performed during four separate beamtimes at the 16 BM-D beamline of the High Pressure Collaborative Access Team (HPCAT, Sector 16) at the Advanced Photon Source at Argonne National Laboratory [21]. A NIST $CeO_2$ standard was used at the beginning of each beamtime to calibrate the sample-to-detector distance, which ranged from 298.5 to 340.9 mm. The x-ray wavelength ranged between 0.3100 and 0.4769 Å. The experimental parameters for all experiments are summarized in Table 1. Table 1 also includes a label for each experiment that will be used to refer to each experiment for the rest of this text. A mar345 image plate detector was used in all experiments except for Ne1 which used a PILATUS3 X CdTe detector. The sample grain sizes were large (up to 20 μm) relative to the X-ray beam spot size (~3x6 μm² full width at the half of maximum intensity). For some experiments, the beam was rastered over the sample in a rectangular grid to collect diffraction from as many differently oriented grains as possible.

NoPTM1 aimed to collect a high density of data points up to 8 GPa to determine the phase transition pressure with high precision. NoPTM2 and NoPTM3 targeted higher pressures to determine the equation of state. NoPTM3 was used to study the final phase after decompression over two days. Ne1 collected quasi-hydrostatic data up to 17 GPa. A single crystal experiment, SC1, was also conducted and provided further verification of the crystal structure. For this experiment, several single crystal grains of $UTe_2$, roughly 20 μm in size, were loaded into the DAC with a 4:1 methanol:ethanol PTM. Data were collected at three pressures up to 5 GPa. At each pressure, the sample was rotated along the vertical axis and perpendicular to the axis of compression in 1° steps from -16° to 16°. A diffraction image was collected during each step for 1 s. Data were collected from the same grain throughout the entire SC1 experiment.

The powder diffraction data were refined using a Le Bail or Rietveld fit and indexed using GSAS-II [22] to determine the volume of the $UTe_2$ sample and the Cu pressure standard.[22] The pressure was determined using the Cu equation of state published by Dewaele *et al.*[23]. The data were fit to the third order Birch-Murnaghan equation of state (EoS),[24] which is given by:

$$P(V) = \frac{3B_0}{2}\left[\left(\frac{V_0}{V}\right)^{\frac{7}{3}} - \left(\frac{V_0}{V}\right)^{\frac{5}{3}}\right]\left\{1 + \frac{3}{4}(B_0' - 4)\left[\left(\frac{V_0}{V}\right)^{\frac{2}{3}} - 1\right]\right\}. \qquad [1]$$

Here, $P$ is the pressure, $V$ is the volume, $B_0$ is the bulk modulus, and $B_0'$ is the pressure derivative of the bulk modulus. A Monte Carlo method described in Ref. [25] with $2\times10^5$ iterations was used to fit the data to Eq. 1. The procedure creates a large number of 'datasets' using the measured P-V coordinates, adds or subtracts a random error within one standard deviation of the measured values, and fits an EoS to each of those datasets. The Monte Carlo Method was chosen to fit the EoS as it accounts for the uncertainty in P and V when fitting the EoS. More detail is provided in the results section. The single crystal data were analyzed using XDS [26] combined with custom software [27].

**Table 1**: Experimental parameters used in each of the five high pressure X-ray experiments on UTe$_2$.

| Exp | Pressure Range (GPa) | Culet diameter (μm) | Sample-to-detector distance (mm) | Wavelength (Å) | Gasket Material | Rastering |
|---|---|---|---|---|---|---|
| NoPTM1 | 0-8 | 400 | 340.9 | 0.4769 | Co$_{66}$Si$_{15}$B$_{14}$Fe$_4$Ni$_1$ | 30x30 μm$^2$ |
| NoPTM2 | 0-25 | 200 | 341.3 | 0.4133 | Co$_{66}$Si$_{15}$B$_{14}$Fe$_4$Ni | 30x30 μm$^2$ |
| NoPTM3 | 0-11 | 200 | 340.9 | 0.4769 | Co$_{66}$Si$_{15}$B$_{14}$Fe$_4$Ni | 30x30 μm$^2$ |
| Ne1 | 1-17 | 400 | 199.7 | 0.4133 | Stainless steel | None |
| SC1 | 0-5 | 200 | 298.5 | 0.3100 | Stainless steel | None |

Because the DAC samples were found to retain the high pressure *I4/mmm* phase after decompression to room pressure, large UTe$_2$ samples (~1 mm$^3$) were compressed in a Paris-Edinburgh (PE) style hydraulic press to enable characterization of a bulk sample after compression and decompression. For this experiment, powdered UTe$_2$ was pressed into a cylindrical pellet 2 mm in diameter and ~0.6 mm thick. The sample was loaded into a PE cell assembly substantially similar to that of Fig. 1 in Ref. [28], but did not utilize ultrasound measurements and included a secondary Al$_2$O$_3$ rod underneath the sample that extended to the bottom anvil to provide a mostly uniaxial compression. The sample was compressed *in situ* at HPCAT beamline 16 BM-B, and energy dispersive XRD from an MgO ring surrounding the sample was used to estimate sample pressure. The sample was held in the PE press at 7-8 GPa overnight, after which it was decompressed and shipped back for postmortem characterization which occurred roughly 2-3 weeks later.

Postmortem characterization involved magnetic susceptibility, specific heat, and electrical resistivity measurements performed on a sample, labeled "high pressure pellet," obtained from the PE press. Ambient pressure powder X-ray diffraction data were collected using a Panalytical Empyrean diffractometer with a Cu K$_\alpha$ source. Magnetic susceptibility measurements were performed in a magnetic field of H=0.1 T from 1.8 K to 350 K using a Quantum Design superconducting quantum interference device (SQUID) Magnetic Properties Measurement System magnetometer. Specific heat measurements from 0.4 to 300 K were carried out in a Quantum Design Physical Properties Measurement System (PPMS) calorimeter that utilizes a quasi-adiabatic thermal relaxation technique. The electrical resistivity was measured in a Quantum Design PPMS cryostat at temperatures 0.35 – 300 K using a standard four-wire technique and a Lakeshore 372 ac bridge.

## Results and discussion

Figure 2 shows a series of X-ray diffraction patterns of polycrystalline UTe$_2$ compressed without a PTM during compression between 0-8 GPa (NoPTM1). Initially the *Immm* phase of UTe$_2$ and the Cu pressure standard are present. A Le Bail fit of a powder diffraction pattern of *Immm* UTe$_2$ is shown in Fig. S1 of the Supplemental Material [28a]. At ambient pressure, the lattice parameters are a=4.168 Å, b=6.130 Å, and

c= 13.959 Å. Reflections that are unique to the *Immm* phase are labeled with blue triangles and the Cu peaks are denoted with black squares. As the pressure is increased from 0 to 5 GPa, the peaks shift to higher values of Q ($Q = \frac{4\pi}{\lambda} \sin\theta$), indicating a decrease in volume with pressure. At 5 GPa, an additional peak, marked by a red circle, is observed at Q=2.49 Å$^{-1}$ and this peak increases in intensity as pressure is further increased. Additional peaks (marked with red circles) also emerge with increasing pressure, while the *Immm* peaks (blue triangles) decrease in intensity. This result points to a mixed-phase region (i.e., both *Immm* and the high-pressure phase are present) occurs between 5-8 GPa. These new peaks correspond to an *I4/mmm* phase of UTe$_2$. These peaks remain present upon decompression to ambient pressure, and were observed to decrease in intensity by ~70% within about 2 days.

A diffraction pattern collected right after decompression to ambient pressure is shown in Fig. 3 (NoPTM3). This pattern was indexed to be a body-centered tetragonal (*I4/mmm*) phase with the CaC$_2$ structure. In this structure, U atoms occupy the 2a (0,0,0) sites and Te atoms sit on the 4e (0, 0, 0.343) sites. The lattice parameters of this *I4/mmm* phase upon decompression to ambient pressure are a= 3.967 Å, and c=10.123 Å. Due to the small sample size, a perfect powder diffraction could not be collected and the modeled intensities with a Rietveld refinement do not perfectly match the measured intensities. Nevertheless, a Rietveld refinement of the pattern indicates the *I4/mmm* phase and elemental Te are present after decompression. The *I4/mmm* phase occupies 10.7% less volume per UTe$_2$ unit than that of the orthorhombic Immm phase at P=0 GPa, meaning that it is significantly denser. Similar phase transitions from the *Immm* to *I4/mmm* space groups have been previously observed or predicted in various materials such as iodine at 43 GPa [29], and XeF$_2$ at 28 GPa [30]. In the phase transition from *Immm* to *I4/mmm* under pressure, the uranium atom site symmetry evolves from C$_{2v}$ to D$_{4h}$. In addition, the distribution of uranium atoms along the orthorhombic c-axis changes from dimer pairs to a uniform chain, resulting in a repeat unit of 10.123 Å along the tetragonal c-axis. Simultaneously, the orthorhombic a- and b-axes undergo compression resulting in the square ab-plane in the final tetragonal structure. The uranium coordination sphere remains eight-fold coordinated during the transformation. In the orthorhombic phase, the uranium coordination is comprised of a distorted square antiprism that transforms to the final square prism in the tetragonal phase upon compression along a and b of the orthorhombic cell.

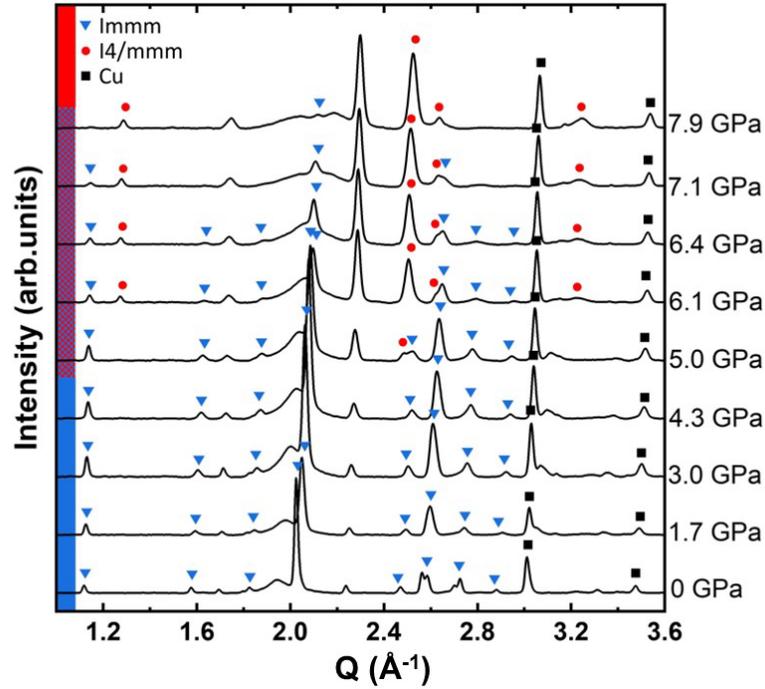

**Figure 2:** X-ray diffraction patterns of UTe$_2$ at pressures between 0 and 8 GPa (NoPTM1). The *Immm* phase is labeled with triangles, the *I4/mmm* phase with circles and the Cu pressure marker with squares. The color bars on the left indicate the phase regions with pressure (blue: *Immm*; purple: mixed-phase; red: *I4/mmm*). Peaks that are common to the *Immm* and *I4/mmm* phases are not labeled. Peaks that are unique to a phase are labeled.

The same phase transition in UTe$_2$ above 5 GPa from *Immm* to *I4/mmm* is also observed in the Ne1 and SC1 experiments, providing further verification of the new structure. Example diffraction patterns from the Ne1 experiment are provided in Fig. S2 [28a]. In the (more) hydrostatic pressure conditions of the Ne1 experiment, the phase transition began at 5.4 GPa and ended at 6.1 GPa. The phase transition is complete in the Ne1 experiment at a lower pressure than in the experiments where no PTM was used. For the Ne1 experiment, a slight deviation from the no PTM (P-V) data for the *I4/mmm* phase is observed. Here the pressure-volume relationship of the *I4/mmm* phase is steeper when the sample is compressed in Ne compared to no PTM, indicating a less stiff bulk modulus. This apparent difference in bulk modulus is typical in studies where compression in quasistatic and non-hydrostatic conditions is used [31, 32]. In the single crystal experiment, the entire sample transformed to the *I4/mmm* phase at a pressure of 4.6 GPa. The lower phase transition pressure observed in the SC1 experiment might be due to kinetic effects as there was more time between each pressure step (40 min) than in the 3-5 min pressure steps in the powder diffraction experiments.

**Figure 3:** Rietveld refinement of the *I4/mmm* phase of UTe$_2$ that was recovered at ambient pressure (NoPTM2), after application of P=10 GPa. The inset shows the structure of *I4/mmm* phase along the (110) direction.

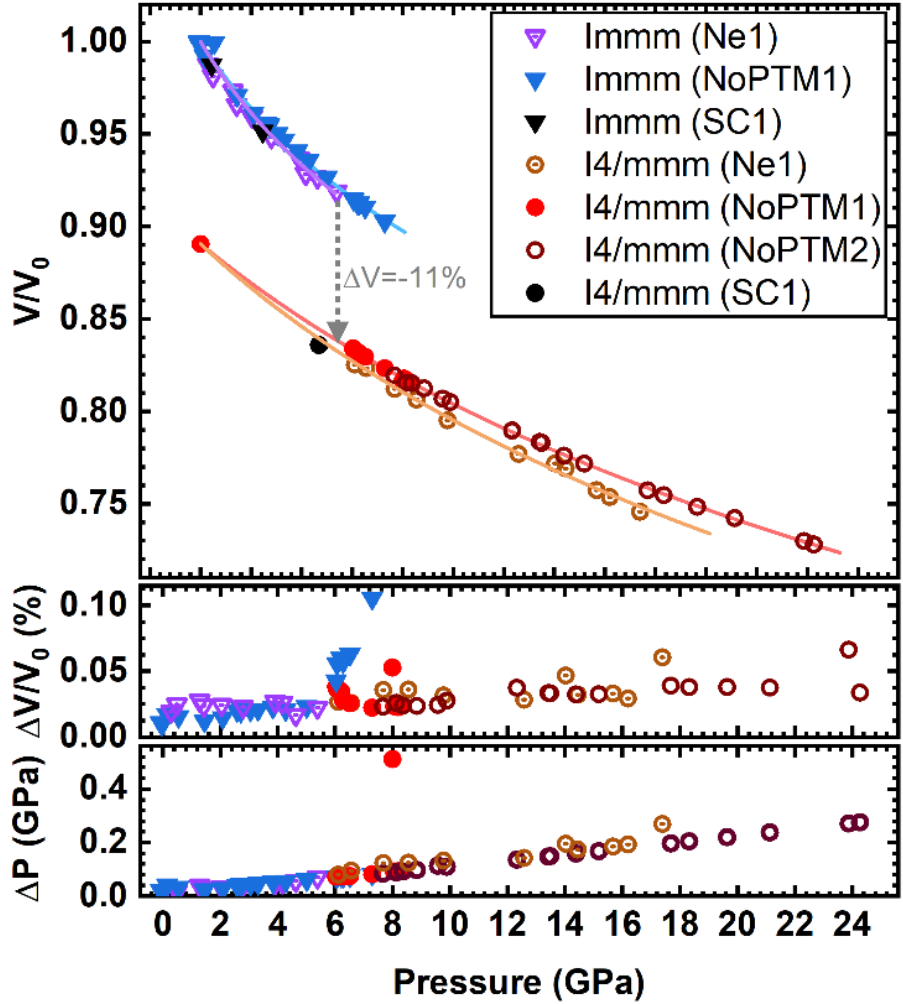

**Figure 4:** The relationship between pressure P and volume V for UTe$_2$ at room temperature. The solid lines show the fits to the third-order Birch-Murnaghan equation of state (Eqn. 1) describe in the main text for the sample where no PTM was used. $\Delta V/V_0$ is the percentage uncertainty in $V/V_0$ from the Rietveld refinement. $\Delta P$ is the uncertainty in pressure determined by using uncertainty propagation to propagate the uncertainty of the volume and the uncertainty in $B_0$ and $B'$ for Cu in Ref. [23].

Figure 4 shows the pressure-volume (P-V) relationship for the *Immm* and *I4/mmm* phases of UTe$_2$ under both quasi-hydrostatic (Ne PTM) and non-hydrostatic conditions (no PTM). In this figure, $V_0 = 357.35$ Å$^3$ is the volume of the *Immm* phase at ambient pressure. The uncertainty on pressure ($\Delta P$) was determined by standard uncertainty propagation. The uncertainties that were propagated to get $\Delta P$ were uncertainty in the volume of Cu from the Rietveld refinement, and the uncertainty in the bulk modulus and its pressure of Cu from Ref. [23]. The data were fit using the Monte Carlo routine described in Ref. [25] with Eqn. [1].

During the fit procedure, $V_0$ was fixed at 357.35 Å$^3$ for the *Immm* phase and 159.11 Å$^3$ for the *I4/mmm* phase. These values were measured at ambient pressure before the experiment (*Immm*) or after decompression (*I4/mmm*). The determined EoS parameters are listed in Table 2:

**Table 2**: Birch-Murnaghan equation of state parameters for UTe$_2$.

| Phase | $B_0$ (GPa) | B' | Pressure medium |
|---|---|---|---|
| *Immm* | 46.0±0.6 | 9.3±0.5 | No PTM |
| *Immm* | 42.5±2.0 | 9.3 (fixed) | neon |
| *I4/mmm* | 78.9±0.5 | 4.2±0.1 | No PTM |
| *I4/mmm* | 70.0±1.1 | 4.1±0.2 | neon |

The values of $B_0$ and B' and their uncertainty are the mean and standard deviation of the values of each variable obtained from the Monte Carlo calculations using the mean and standard deviation functions in Matlab, as shown in Fig. S3 [28a]. Because the Ne PTM data for the *Immm* phase only contained ten data points with considerable scatter, B' was fixed at 9.3, the value for the no PTM measurement, for this least-squares fit. A comparison with the fit with B'=9.3 GPa and the Monte Carlo fit is found in Fig. S4 [28a]. For further characterization of the structural evolution of UTe$_2$ with pressure, the value of $c/\sqrt{a^2+b^2}$ for the *Immm* phase and c/a ratio for the *I4/mmm* phase are provided in Fig. S5 [28a]. For the *Immm* phase, $c/\sqrt{a^2+b^2}$ initially decreases until ~2 GPa and then increases until the phase transition pressure. For the *I4/mmm* phase, c/a is found to decrease with pressure from 2.54 at 6 GPa to 2.51 at 20 GPa. A change in the c/a or $/\sqrt{a^2+b^2}$ with pressure generally suggests a non-isotropic compressibility. However, in the case of using no PT, preferred orientation caused by plastic deformation of the sample may also impact c/a or $/\sqrt{a^2+b^2}$.

Figure 5 provides a summary of results on a pellet of UTe$_2$ after being held at ~7-8 GPa in a PE press overnight followed by decompression to ambient pressure (labeled "high pressure pellet"). The characterization measurements were conducted 2-3 weeks after decompression. Powder x-ray diffraction results indicate that the UTe$_2$ pellet contains a mixture of UTe$_2$ in the orthorhombic *Immm* phase and free tellurium. The magnetic susceptibility χ(T) of the pellet at H=0.1 T, scaled by a factor of 1.1 to account for the volume fraction (~90%) of orthorhombic UTe$_2$ present in the sample, is shown in Fig. 5a. The shape of χ(T) of the pellet closely resembles the polycrystalline average of the magnetic susceptibility ($\chi_{poly}$ = ($\chi_a$+ $\chi_b$+ $\chi_c$)/3) of a single crystal of UTe$_2$. The inset to Fig. 5a shows the strong increase of χ(T) below 50 K present in both the high pressure pellet and $\chi_{poly}$, which is dominated by $\chi_a$ below about 50 K [11]. At high temperature, the magnetic susceptibility of the high pressure pellet follows a modified Curie-Weiss law, χ(T) =C/(T-θ$_{CW}$) + χ$_0$, with an effective moment μ$_{eff}$=3.2 μ$_B$/U atom, Curie-Weiss temperature θ$_{CW}$= -180 K, and a temperature-independent offset χ$_0$=0.0001 emu/mol. At low magnetic field (H=0.0005 T), zero-field-cooled/field-cooled ZFC/FC measurements of χ(T) revealed a tiny difference between the ZFC and FC signal (not shown), at most 0.005% of 4πχ, indicating the high pressure pellet exhibits little (if any) bulk superconductivity at 2 K. The specific heat, plotted as C/T, of the UTe$_2$ high pressure pellet (normalized by a factor of 1.1 corresponding to about 90% UTe$_2$, black symbols) and the unpressurized single crystal (blue symbols) is displayed in Fig. 5b. While the single crystal shows a sharp superconducting transition at T$_c$=2.0 K, the high pressure pellet only exhibits a very small change in slope below ~1.5 K, again indicating little, if any, superconductivity. No evidence for superconductivity above 0.35 K was found via electrical resistivity measurements. Superconductivity in $UTe_2$ is known to be sensitive to defects, impurities,[33] subtle changes in lattice parameters, and other structural parameters such as anisotropic displacement parameters [11,34]. Furthermore, samples grown by a sub-optimal chemical vapor transport method results in non-superconducting crystals with uranium vacancies and smaller lattice volume [11, 35-37]. Therefore, a sensible scenario is that superconductivity is suppressed in orthorhombic $UTe_2$ subjected to 8 GPa under non-hydrostatic conditions due to a combination of strain, defects, and small changes in lattice parameters.

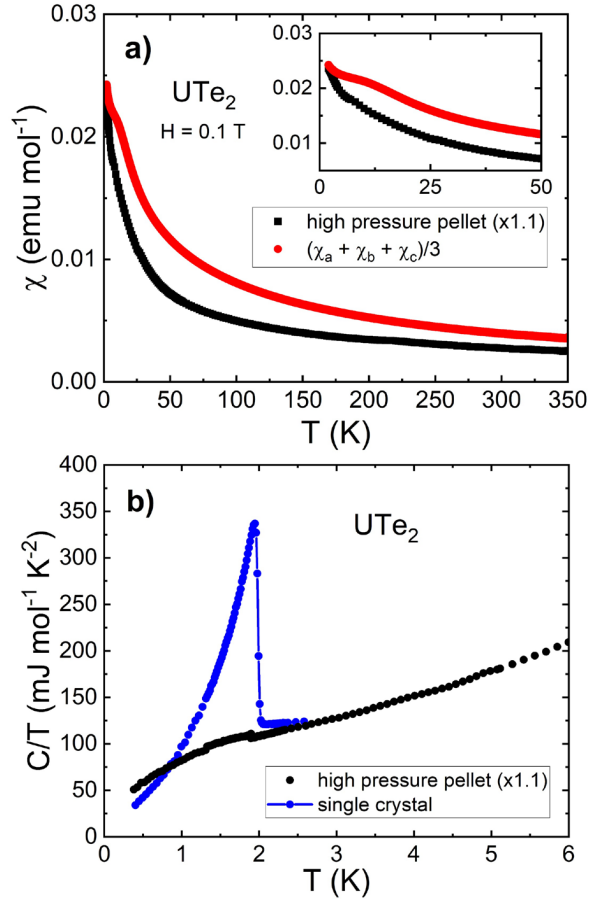

**Figure 5: a)** Magnetic susceptibility $\chi(T)$ of a high pressure pellet (black symbols), normalized by a factor of 1.1 (see text for details) and polycrystalline average of $\chi_{poly}$ of a single crystal of $UTe_2$ (red symbols). Inset: $\chi(T)$ of a high pressure pellet (black symbols) and $\chi_{poly}$ of a single crystal below 50 K. **b)** Specific heat, $C/T$ vs $T$, of a high pressure pellet, normalized by a factor of 1.1 (see text for details), and a single crystal of $UTe_2$.

**Discussion**

Trends in the high-pressure behavior of uranium-based materials may provide insight into the mechanism for the *Immm* to *I4/mmm* phase transition in UTe$_2$ [38]. Within the actinide elements, low-symmetry structures occur for the light actinides (e.g., orthorhombic α-U), where the 5f-electrons are itinerant and participate in bonding. Essentially, when the f-electron wavefunctions overlap and form bands, they imprint their low symmetry on the structure [39]. For the heavy actinides (Am and beyond) where the 5f-electrons are further apart and localized, the actinide elements crystallize in high-symmetry, hexagonal close packed structures. The basic situation of itineracy/localization of the 5f electrons and bonding has been captured in a Hill plot of ordering temperature (superconducting or magnetic) vs shortest U-U distance $d_{U-U}$ [40]. Here, Hill pointed out that in U-based compounds with $d_{U-U}$<3.5 A, superconductivity or Pauli paramagnetic ground states occurred, while in U compounds with $d_{U-U}$>3.5 A, magnetic order tended to be present. This analysis neglected f-electron/conduction electron hybridization. The periodic Anderson model (PAM) [41] provides an accurate description of f-electron compounds, including Coulomb repulsion (U) and hybridization ($V_{kf}$) between f-electrons and conduction electrons, though it has only been solved in limiting cases, e.g., the Kondo model [42]. Within this modern viewpoint of the PAM, complex f-electron behavior and emergent ground states (e.g., magnetic order, superconductivity, intermediate valence) arise from the interplay between Coulomb repulsion, which tends to localize the f-electrons, and hybridization, which promotes f-electron itinerancy; here, the f-electron behavior in U-based compounds spans the range between fully itinerant and fully localized, as is observed experimentally.

The higher symmetry of the high-pressure *I4/mmm* phase in UTe$_2$ suggests that the phase transition observed in UTe$_2$ at 5 GPa is driven by increased 5f-electron localization with pressure within a mixed valence ground state of UTe$_2$ at ambient pressure. An intermediate-valence state between 5f$^2$ (U$^{4+}$) and 5f$^3$ (U$^{3+}$) configurations in UTe$_2$ at ambient pressure has been observed using X-ray absorption near-edge spectroscopy (XANES) [15] and 4f core-level photoemission spectroscopy [25]. Electrical resistivity, thermal expansion, and scanning tunnelling microscopy measurements indicate a characteristic (Kondo) energy scale for the intermediate valence state of $T_K$~30 K [43]. The value of the Gruneisen parameter, $\Gamma=(B_0/T_K)(dT_K/dP)$ = -30 [3], consistent with the decrease in $T_K$ with applied pressure [15, 33], is typical of intermediate valence materials [44]; though, in general, the sign of the Gruneisen parameter for U-based compounds is positive, similar to Ce-based heavy fermion/intermediate valent materials [44]. The intermediate valence ground state for UTe$_2$ is also in agreement with the moderate effective mass enhancement m*~30 m$_e$ [3, 10], where m$_e$ is the bare electron mass. When the pressure is increased above 1.4 GPa, magnetism emerges in UTe$_2$ and is accompanied by an increase in the contribution from the U$^{4+}$ (5$f^2$) valence state is observed via XANES, indicating the 5f electrons become more localized with pressure [15], The phase transformation in UTe$_2$ at 5 GPa from a low symmetry, orthorhombic structure (*Immm*) to a higher symmetry, tetragonal structure (*I4/mmm*) indicates further localization of (some of) its 5f electrons. Indeed, a significant increase in the smallest U-U distance ($d_{U-U}$) at 5 GPa and a large increase in bulk modulus (Table 2) is found between the two structures: $d_{U-U}$ = 3.6 Å in the *Immm* structure, while $d_{U-U}$ = 3.9 Å in the *I4/mmm* phase; however, there is also a large volume decrease at 5 GPa of $\Delta V/V_0$~11% from the orthorhombic to tetragonal phase. Similar trends in 5f-electron localization are found in the UM (M=C, N, P, S) compounds, in which the 5f electrons are itinerant (small U-U distances $d_{U-U}$<3.5 A), and undergo a structural phase transformation from higher to lower symmetry structures (i.e., UN phase transforms from cubic to rhombohedral at 29 GPa [37, 44]). For the UX compounds with X=Se, Te, in which the 5f electrons are more localized (large U-U distances), phase transitions under pressure occur from lower symmetry to higher symmetry structures (i.e., NaCl to CsCl structure) [37]. For example, in UTe, a phase transition from the NaCl phase to the CsCl structure occurs at 11 GPa [46-48] with a significant increase in bulk modulus (44.6 GPa in the NaCl phase vs. 60.1 GPa in the CsCl phase) [46]. Furthermore, a large volume decrease ($\Delta V/V_0 \sim 9\%$) is associated with the phase transition at 11 GPa; after decompression, the high-pressure CsCl phase of UTe remains present to low pressures (0.5 GPa), similar to UTe$_2$ [45-47].

A framework for understanding the increase in 5f-electron localization with pressure in UM$_2$Si$_2$ (M=Fe, Ru, Pd, Ni) compounds, which also crystallize in the *I4/mmm* space group has been proposed by Amorese *et al.* [49]. In this case, an increase of 5f$^2$ (U$^{4+}$) weight with increasing pressure in the mixed-valence (5f$^3$ / 5f$^2$) ground state decreases the magnetic exchange interaction between the 5f-electrons and the conduction electrons. The decrease in the magnetic exchange with P decreases the Kondo interaction. This decrease in magnetic exchange with pressure (or, alternatively, unit cell volume) is consistent with the evolution of the ground state properties from Pauli paramagnetic (UFe$_2$Si$_2$), to superconducting (URu$_2$Si$_2$), to magnetically ordered (UNi$_2$Si$_2$, UPd$_2$Si$_2$), and the magnitude of the ordered moments for M=Ni, Pd. UTe$_2$ fulfills a number of criteria of the framework above: an initial mixed-valence ground state between 5f$^2$ and 5f$^3$ configurations and an increase in localization of its 5f electrons with increasing pressure. Thus, if this scenario applies to UTe$_2$, it is a significant decrease of the 5f$^2$ weight (i.e., more 5f electron localization) and is expected at the phase transition at 5 GPa. In addition, theoretical Periodic Anderson Model calculations of UTe$_2$ indicate a change in the momentum-dependent magnetic susceptibility with pressure [16]. For small pressure (delineated in units of an enhancement factor of hopping integrals), ferromagnetic (q=0) fluctuations dominate the behavior of UTe$_2$, while at high pressure, antiferromagnetic (finite q) fluctuations dominate; this change in the nature of the magnetic fluctuations with increasing pressure is driven by a change in orbital character [16]. Our X-ray diffraction results provide information to put these PAM calculations on an absolute pressure scale to better compare them to experiment. Various thermodynamic measurements suggest antiferromagnetically ordered states above 1.4 GPa (Fig. 1b). A recent NMR study in UTe$_2$ at 1.8 GPa of site-specific $^{125}$Te relaxation rate measurements [50] found signatures of nearly-commensurate antiferromagnetic spin fluctuations that grow with decreasing temperature towards the long-range magnetic ordered state ($T_{m1} \sim 8$ K), in contrast to more uniform (possibly ferromagnetic) fluctuations observed at ambient pressure above the superconducting transition [51]. Clearly, further measurements are needed to understand the evolution of the behavior of UTe$_2$ under pressure, particularly in the high pressure tetragonal phase above 5 GPa [52].

**Conclusions**

In this study, the crystal structure and equation of state of UTe$_2$ was determined up to 25 GPa. A phase transformation from the orthorhombic *Immm* to the tetragonal *I4/mmm* structure was observed at 5 GPa under quasi-hydrostatic (Ne) conditions, while it occurred between 5 and 8 GPa under non-hydrostatic conditions (no pressure-transmitting medium). The *I4/mmm* phase remained present after decompression to 0 GPa for about 2 days at room temperature. No superconductivity is found above 0.35 K when UTe$_2$ transforms back to the orthorhombic *Immm* phase after about 2-3 weeks. The UTe$_2$ bulk modulus and its pressure derivative were determined to be: B$_0$=46.0±0.6 GPa, B'=9.3±0.5 (no pressure medium) and B$_0$=42.5±2 GPa, B'=9.3 (neon) for the *Immm* phase. For the *I4/mmm* phase, the equation of state parameters were found to be B$_0$=78.9±0.5 GPa and B'=4.2±0.1 (no pressure medium), and B$_0$=70.0±1.1 GPa and B'=4.1±0.2 (neon). The phase transition at 5 GPa is likely driven by the increase in localization of the 5f electrons of UTe$_2$ with pressure.

**Acknowledgements**

L.Q.H. was supported by the Laboratory Directed Research and Development (LDRD) program of Los Alamos National Laboratory (LANL) and the G. T. Seaborg Institute. A.W. and S.M.T were supported by the LDRD program at LANL. E.D.B., P.F.S.R, and J.D.T. were supported by the U.S. Department of Energy (DOE), Office of Basic Energy Sciences, Division of Materials Science and Engineering under the "Quantum Fluctuations in Narrow-Band Systems" project. LANL is operated by Triad National Security, LLC for the DOE-NNSA under Contract No 89233218CNA000001. Portions of this work were performed at HPCAT (Sector16), Advanced Photon Source (APS), Argonne National Laboratory. HPCAT operations are supported by DOE-NNSA's Office of Experimental Sciences. The Advanced Photon Source is a U.S. Department of Energy (DOE) Office of Science User Facility operated by Argonne National Laboratory under Contract No. DE-AC02-06CH11357.

# Supplemental Material: Metastable phase of UTe$_2$ formed under high pressure above 5 GPa

L. Q. Huston[1], D. Y. Popov[2], A. Weiland[1], M. M. Bordelon[1], P. F. S. Rosa[1], R. L. Rowland II[1], B. L. Scott[1], G. Shen[2], C. Park[2], E. K. Moss[1], S. M. Thomas[1], J. D. Thompson[1], B. T. Sturtevant[1], E. D. Bauer[1,a]

[1]Los Alamos National Laboratory, Los Alamos, NM 87545, USA

[2]HPCAT, X-ray Sciences Division, Argonne National Laboratory, Lemont, IL, 60439, USA

a) edbauer@lanl.gov

**Supplemental Figures**

Figure S1 contains a Le Bail fit of a diffraction pattern containing *Immm* UTe$_2$ at 0 GPa. The fit yields the lattice parameters a=4.168 Å, b=6.130 Å, and c=13.959 Å, which are in good agreement with previous studies [S1, S2].

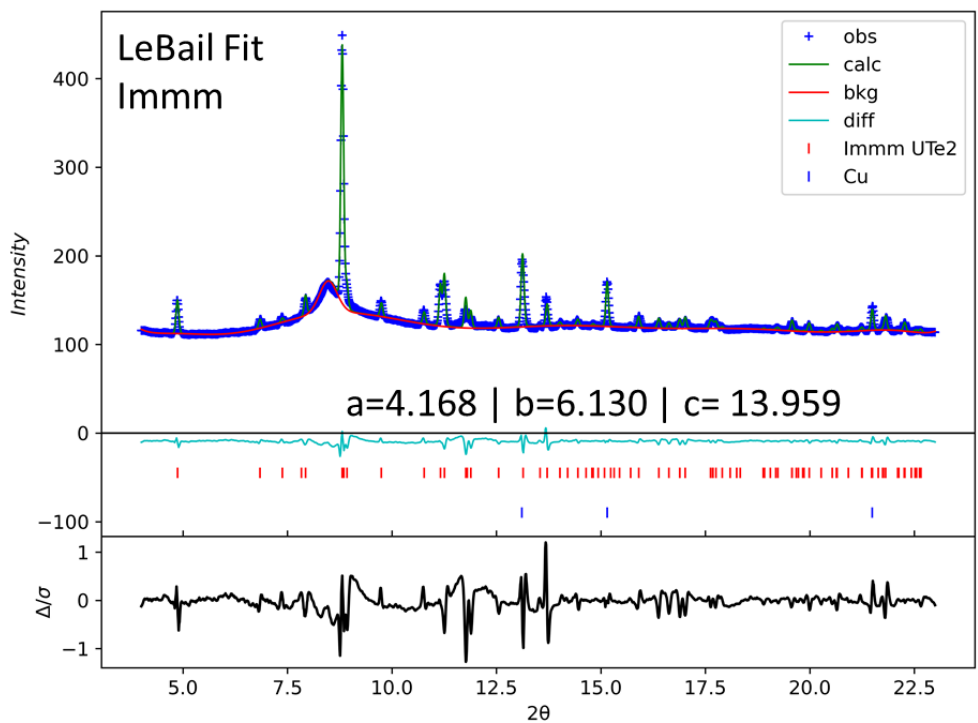

**Figure S1:** Le Bail fit of the *Immm* phase of UTe$_2$ at ambient pressure. The lattice paramteres are given in angstroms.

Figure S2 shows x-ray diffraction patterns of UTe$_2$ that was compresed in a Ne pressure transmitting medium. Figure S2(a) show an x-ray diffraction image of UTe$_2$ in a Ne pressure transmitting medium at 0.1 GPa. The red regions of the image that were masked during integration. These regions correspond to either the gaps in the detector or the diffraction spots from the single crystal sapphire windows, which were used as secondary containment for the sample. The UTe$_2$ pattern has a "spotty" appearance, which comes from only a few single crystalline grains of UTe$_2$ within the X-ray beam of ~3-5 microns in diameter. Figure

S2(b) shows diffraction patterns obtained from integration of the x-ray diffraction image. At 0.1 GPa, UTe$_2$ exists in the *Immm* phase. Peaks of this phase are marked with blue triangles. Note that due to the beam hitting a slightly different region of the sample at each pressure, the intensity of each peak varies and differs between diffraction patterns. UTe$_2$ remains in the *Immm* phase at 2.7 GPa and 4.6 GPa. At 6.1 GPa, a decrease in the intensity of the *Immm* peaks occurs and new peaks, associated with the tetragonal (*I4/mmm*) phase appears (red circles). At 14.4 GPa, the sample is entirely in the *I4/mmm* phase.

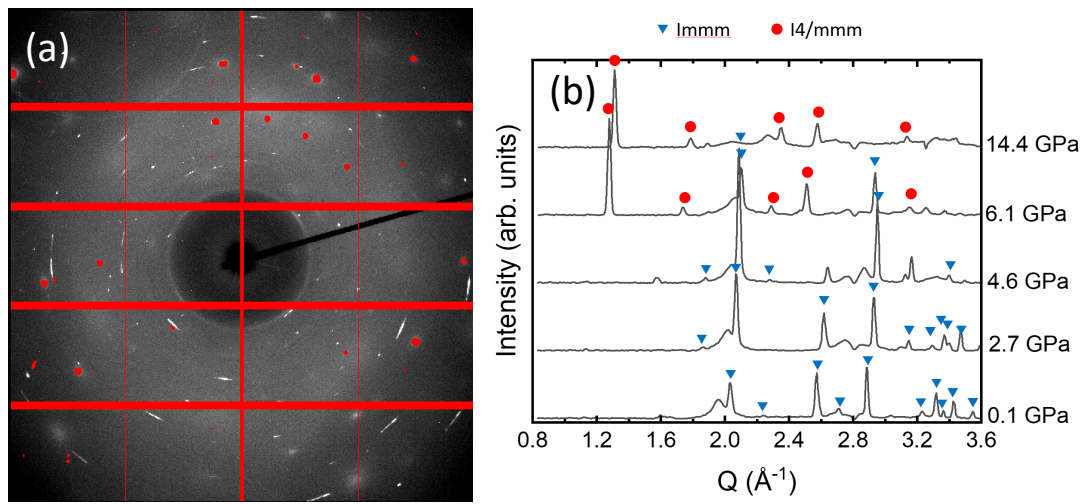

**Figure S2:** (a) An example of an x-ray diffraction image collected of *Immm* UTe$_2$ compressed in a Ne pressure transmitting medium at 0.1 GPa. (b) Selected integrated diffraction patterns taken of UTe$_2$ compressed in a Ne presume medium at various pressures.

Figure S3 shows the values of $B_0$ and $B'$ determined using the Monte Carlo method described in Ref. [S3]. Each plot consists of $2\times10^5$ data points.

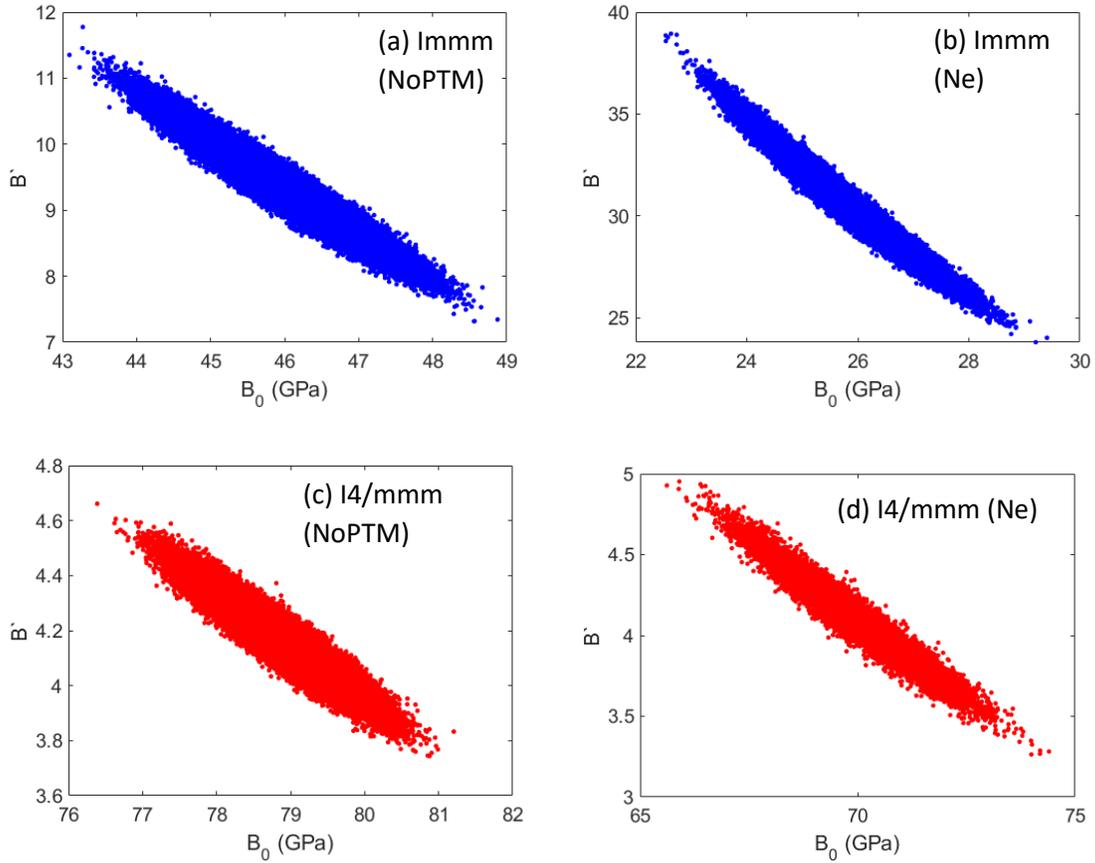

**Figure S3:** Results for the Monte Carlo fitting of the experimental data for the (a) *Immm* (NoPTM), (b) *Immm* (Ne), (c) *I4/mmm* (NoPTM), and (d) *I4/mmm* (Ne) phases of UTe$_2$.

Figure S4 shows the fit of the *Immm* phase of UTe$_2$ compressed in a neon pressure transmitting medium to the Birch-Murnaghan equation of state. The light purple line shows the equation of state determined using the Monte Carlo Method described in Ref. [S3] with both $B_0$ and $B'$ free. This fit yielded $B_0$=25.8±0.7 GPa, $B'$=30.5±1.6, which is likely unphysical. As only 10 data points for the *Immm* phase in Ne were collected, a more reasonable least-squares fit (dark grey) is obtained where the pressure derivative of the bulk modulus, $B'$, was fixed at 9.3 GPa, the value from the no pressure transmitting medium fit. This fit gives a value of $B_0$= 42.5±2.0 GPa.

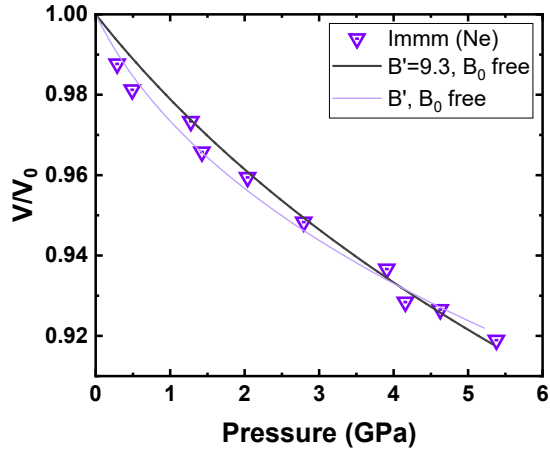

**Figure S4:** The pressure-volume relationship of *Immm* UTe$_2$. The solid lines show the fits to the Birch-Murnaghan equation of state with B$_0$ and B' free (purple line), and B' fixed at 9.3 (black line).

Figure S5 shows the evolution of $c/\sqrt{a^2+b^2}$ for the *Immm* phase and c/a for the *I4/mmm* varies with pressure. For the *Immm* phase, $c/\sqrt{a^2+b^2}$ initially decreases until ~2 GPa and then increases until the phase transition pressure. For the *I4/mmm* phase, c/a decreases with pressure from 2.54 at 6 GPa to 2.51 at 20 GPa.

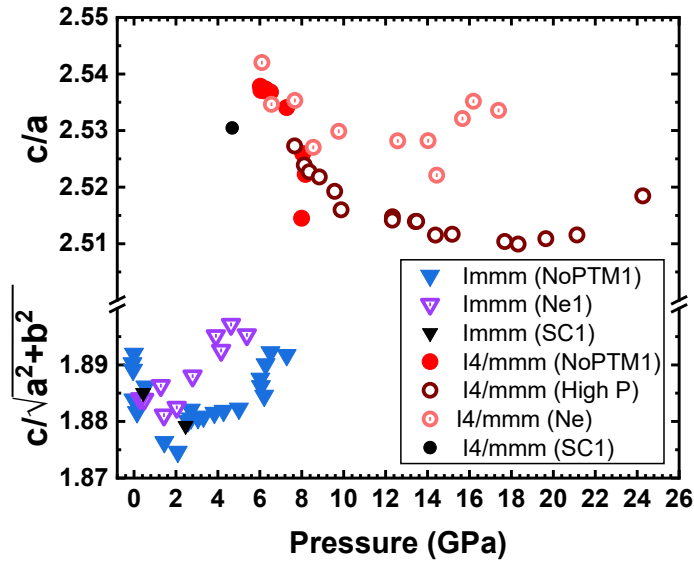

**Figure S5:** The relationship between pressure and the $c/\sqrt{a^2+b^2}$ ratio for the *Immm* phase and the c/a ratio for the *I4/mmm* phase of UTe$_2$ compressed without a pressure transmitting medium.